# FEM modeling and simulation of broadband ultrasonic transducers with randomized inhomogeneous backing material

Eduardo Moreno[1], Wagner C. A. Pereira[2], Marco Antonio von Kruger[2], Lorenzo Leija[3] and Antonio Ramos[4]

(1) Ultrasonic Group, ICIMAF, Calle 15 No 551 Vedado, La Habana, Cuba. moreno@icimaf.cu
(2) Biomedical Engineering Program, COPPE/UFRJ, 2030 Horacio Macedo Ave, 21941-914, Rio de Janeiro, Brazil. wagner@peb.ufrj.br
(3) Department of Electrical Engineering, Bioelectronics Section, Centro de Investigación y de Estudios Avanzados del Instituto Politécnico Nacional, Cinvestav, *México* City, México. lleija@cinvestav.mx
(4) R&D Group "Sistemas y Tecnologías Ultrasónicas", Instituto de Tecnologías Físicas y de la Información, CSIC, Madrid, Spain. antonio.ramos@csic.es

**Abstract.**

A FEM application for the accurate design of composite backing of ultrasonic transducers is presented. The idea is to obtain the dependence between the volume ratio of the tungsten powder in an epoxy matrix used as a backing and the final pulsed RF signal emitted by the backed transducer. An inhomogeneous material model for the backing was adopted instead of the classic complex geometrical FEM. The model was developed in Comsol 5.5 using two of its physics interfaces. The first based on classic FEM or continuous Galerkin for ultrasonic propagation in the backing and the second based on a discontinuous Galerkin option for fluid pulsed propagation. In this way it is possible to avoid the use of an impedance-based model. The calculations for the resulting bandwidth are good enough as predicted for Desilets for the 1D case by using the KLM circuital model. We worked in a 2D space, where the transducer face vibration is far away from the assumed piston-like considered by the unidimensional KLM model.

*Introduction*

In ultrasonic piezoelectric transducers for flaw detection and imaging, the backing component is an important element in order to increase the desired broadband of the E/R ultrasonic pulses. Metal-loaded backings have been the most commonly used for this purpose. These materials use metal powder suspended in epoxy resin where the acoustic properties could be controlled according to the powder concentration. During many years, several analytical theoretical models have been proposed in order to explain the acoustic impedance obtained in this composite material. However, they have limitations according to the scattering phenomena presented.

The objective of this paper is to present a FEM model using inhomogeneous material based on the statistical distribution of tungsten powder in epoxy resin. Additionally, our FEM model considers the analysis of the bandwidth of the ultrasonic pulse emitted in fluid media, as a merit figure of the performance of the final ultrasonic transducer. Then we do not precise to evaluate the acoustic impedance of the backing as in previous FEM and analytical models developed.

*Theoretical Considerations*

Many pulsed ultrasonic applications use a piezoelectric transducer with active elements such as lead zirconate titanate (PZT). This active element has a relative low vibration absorption; then pulses with many cycles are produced and introduced to the inspected medium through the front acoustic port or load face. In many industrial and medical applications, the number of cycles must be quite limited for increasing the axial resolution. This can be achieved by using a backing block in the back acoustic port with an acoustic impedance close to that of the piezoelectric material and having in addition a high ultrasonic attenuation [1].

Using the delta reflection model [2], the transmitted amplitude entering into the backing for normal incidence is given by:

$$T = \frac{2Z_B Z_P}{(Z_B + Z_p)} \tag{1}$$

Where $Z = \rho v$, is the acoustic impedance and $\rho$ and $v$ are the density and longitudinal ultrasonic velocity for each material respectively. The subscript $B$ and $p$ are for the backing and the piezoelectric element respectively. With an air backing, for instance, the value of $T$ is close to zero. The reflected amplitude given by $R = (Z_P - Z_B)/(Z_P + Z_B)$ has a maximum value. Then practically all the amplitude is reflected inside the piezoelement and consequently many cycles are emitted from opposite front face [2]. Hence it is increasing the pulse length and diminishing the axial resolution. On the other side, when the acoustic impedance of the backing increases, the number of cycles of the ultrasonic pulses begins to decrease [2].

Nevertheless, equation (1) is not good enough to express the optimum bandwidth that can be obtained for a given acoustic impedance of the backing. Desilet et al., using the KLM circuit [3], established this requirement: that the mechanical and electrical quality factor $Q$ should be equal to each other for optimum bandwidth, which yielded to the following relationship [3]:

$$\frac{Z_B + Z_L}{Z_P} = k_t \sqrt{2} \tag{2}$$

Where $Z_L$ is the acoustic impedance of the front acoustic load and $k_t$ is the thickness coupling factor [4]. As expressed before an optimum bandwidth is the final objective of improving the axial resolution.

For medical applications, the value of $Z_L$ could be neglected and assuming typical value of $k_t$ of 0.5 for PZT elements, equation (2) could be expressed as:

$$Z_B \approx 0.7 Z_P \tag{3}$$

Expression (3) shows that, for this medical case, it is not necessary to obtain a backing material with an acoustic impedance equal to the piezo-element impedance, in order to obtain a maximum bandwidth.

Many backing layers consist of composites made from epoxy resin with a high ultrasonic absorption, filled with tungsten powder. The epoxy resin, although presenting high ultrasound attenuation, has a low density of approximately 1140 kg/m³ relative to piezoceramics (around 7500 kg/m³), and then is not good enough as a backing. The use of tungsten powder (with a high density around 17800 kg/m³) added to epoxy resin, allows compensating the low density of the resin itself. This is the main reason why tungsten is commonly used as a filler backing material. The use of different volume fractions of Tungsten/epoxy-resin (VFTE) allows obtaining composite backings with several acoustic impedances $Z_B$ enabling the fabrication of transducers with different bandwidths [2].

The composite, made with particles of tungsten embedded in an epoxy resin matrix, presents a connectivity that varies according to the VFTE relations. Using the Newmann definitions of composites [5], it is expected for lower VFTE values, a 0-3 connectivity (there is no connection between tungsten particles and connections in all dimensions for epoxy resin). For higher VFTE values, it is expected tungsten cluster formations and the connectivity passes to a 3-3 condition where both, epoxy resin and tungsten, make connections in all dimensions. This composite condition has been classified as 0-3/3-3 connectivity [6,7].

From the acoustic impedance values of the backing material, obtained with several VFTE volume fractions, we can predict the ultrasonic signal shape and then the bandwidth. For instance, it is possible to use the KLM model [8] for this purpose. The problem is that this model is unidimensional; then novel models using Finite Element Method (in this work, Comsol Multiphysics® Software) improve these limitations [9,10]. However, many of these previous models still assume a single value of the acoustic impedance of the backing (or averaging approach) that depends on the VFTE.

The average acoustic impedance value of the composite and its volume fraction dependences is not so obvious. Hence, several theoretical models have been proposed. These models are more specific to the velocity component (elastic constant) of the acoustic impedance than for the density $\rho$ itself. For this second parameter, the backing composite density is given by [1, 11]:

$$\rho_B = \rho_T v_T + \rho_E v_E \qquad (4)$$

Where subscript $T$ and $E$ corresponds to both phases (tungsten and epoxy resin), $v$ is the volume fraction of each phase, and $v_T + v_E = 1$. In our case $v_T$=VFTE. Then equation (4) could be expressed as:

$$VFTE = v_T = \frac{\rho_B - \rho_E}{\rho_T - \rho_E} \qquad (5)$$

With equation (5) it is possible to calculate the VFTE from the value of $\rho_B$. This last one could be evaluated, for instance, from the average density values obtained from FEM models.

For the case of the other acoustic impedance component: the composite ultrasonic velocity and its relation with the VFTE, several models have been studied specifically for the elastic properties (as a velocity component) by some kind of averaging. Some of these are the classic Reuss [12] and the Voigt models [13], based on assumptions concerning geometry and physical elastic behavior of the two-phase components. More recently, Hashin and Shtrikman [14] proposed new models using elastic energetic considerations where upper and lower bound-limits for the effective properties are obtained. These bound-limits show the range of the velocity values for each VFTE fraction [15]. Nevertheless, these bound limits affect the accuracy of transducer design

Other approaches, in order to obtain more accurate predictions, were made using scattering models [16]. However, all these models (like the previous explained above) are approximations with a great influence of the VFTE range. In this case, for low VFTE values, a single scattering model could be assumed, which it is not possible for high VFTE values, where multiple scattering is expected. The use of finite element method or FEM could be a good approach as in Ref [7]. In the case of piezoelectric composite (that could be considered similar to our composite backing), Ref [17] presents a method that consists in the development of numerical models composed of a unit cell with piezoelectric fibers made of PZT embedded in a non-piezoelectric matrix (epoxy resin).

In classics developed FEM models, for composite backing material the tungsten particles are located by random geometry coordinates, which means a complex FEM geometry with two materials presented. A first approach strategy is to mesh the epoxy matrix and the random particles with the objective of obtaining the ultrasonic velocity for each VFTE value. As the particle size is small, a great number of degrees of freedom are generated, because the mesh arrangement must include boundary conditions between the tungsten particle and epoxy matrix. In addition, the results of FEM simulations show a distortion of the wavefront that is necessary to consider the velocity evaluation into the composite backing. This distortion is more affected for high VFTE values where clusters of tungsten powder are formed. It could be concluded that in this first FEM approach strategy is not good enough. The mesh size and time interval are also quite critical and computer time and efficiency could be affected.

There is another second FEM approach strategy using Comsol. The idea is to develop a composite backing model based on only one inhomogeneous material with a single geometrical domain. This is a completely different approach from the previous ones, which uses two materials in a complex geometric domain. Additionally, this single composite backing domain could be combined with the piezoelectric element to form a transducer first FEM model. The vibration obtained from the face of the transducer, under voltage pulse excitation, could be used as the input signal in a water medium for a second FEM model. Then, it is possible to evaluate the bandwidth of the transducer with the RF signal emitted in water, without the knowledge of the backing acoustic impedance. This is the fundamental objective of this paper, where the central frequency and bandwidth are calculated for several VFTE conditions using this new FEM model. The novel mathematical tools developed, like discontinuous Galerkin, for the last Comsol versions make this possible.

*MATERIAL AND METHODS*

The simulation was entirely implemented in the COMSOL Multiphysics® platform, version 5.5, including the use of its library of material properties.

*Finite Element model*

As a first step, a 2D dimensional model was adopted in the Model Builder in Comsol 5.5. This model was split in two models with their own individual Time-Dependent-STUDY, respectively. The first one was a multiphysic model, named Transducer Model, for the piezoelectric transducer (2 MHz piezoceramic plus backing) including a water layer. This model uses conventional FEM or continuous Galerkin method (CG) and it is formed by SOLID MECHANICS, ELECTROSTATIC and PRESSURE ACOUSTIC TRANSIENT from Comsol Physics. The ultrasonic pulse propagation is the second model developed by the discontinuous Galerkin method (DG) and it only uses the PRESSURE ACOUSTIC TIME EXPLICIT (named Propagation Model).

Up to this moment it is not possible to use DG with piezoelectric materials in Comsol 5.5. Then these two models CG and DG are necessary and used in a concatenated form. The first for the transducer itself with a backing and the second for the pulse propagation in water. Each of them uses specific time-depended solvers that are not compatible to each other [18]. The use of DG model is very convenient for pulse propagation according to its advantage of a relative coarser mesh size around 1.5λ. This makes possible to simulate larger areas of wave propagation, with fewer degrees of freedom. On the other side, the classic FEM (CG) is implemented for the piezoelectric transducer (piezoceramics plus backing) with a relatively small layer of fluid. This fluid layer is necessary to make an indirect coupling between these two models as described in [18].

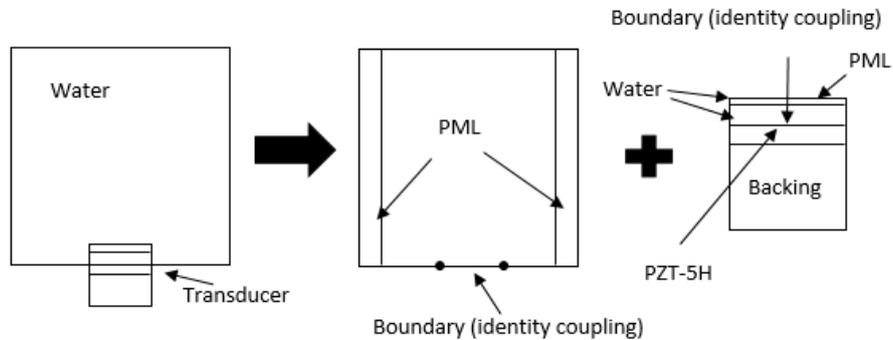

Fig. 1. Description of the physical domain (left) split in two domains: center for fluid medium for DG, right transducer with a backing, piezoceramics PZT-5H and fluid layer (for CG). PML = Perfect Matching Layer. The piezoelectric PZT-5H has a frequency $f_0$ = 2 MHz.

Fig. 1 shows (left) a diagram of the general physical domain of a medium with a transducer. This domain is divided in two sub-domains for each model: Transducer and Propagation Model. The connection between them is made through a boundary *identity coupling* defined in each model. In the Transducer Model, this coupling is defined specifically in the internal boundary between the piezo-element and the fluid layer (Fig. 1, right). In the Propagation model, the connection coupling is defined on the boundary as shown in the center of Fig. 1.

For the Propagation model, the geometric domain is a square of 40x40 mm with 5 mm of PML layer on both sides (Fig. 1, center). Fig. 2 display details of the Transducer model, where the wavelength λ=0.75 mm, assuming fluid velocity $C_0$= 1500 m/s and a frequency $f_0$=2 MHz. This frequency corresponds to the thickness mode of piezo-element PZT-5H with 1 mm of thickness.

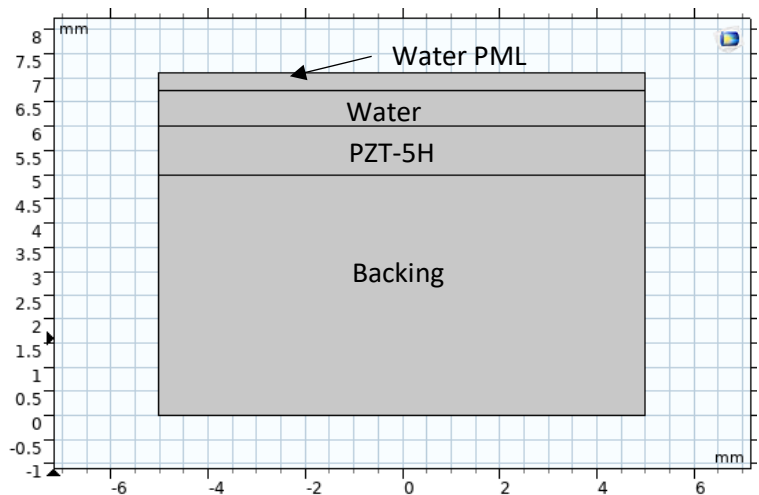

Fig. 2. Transducer geometrical domain used in Transduce Model. Backing 10x5 mm. PZT-5H 10x1 mm. Water layer 10xλ mm. PML 10xλ/2 mm.

The Propagation model domain determines the time-dependent interval computed. In this case, assuming a water medium with ultrasonic velocity 1500 m/s, the propagation was evaluated from 0 to 40[mm]/1500[m/s]=2.667 µseg. The time-step was $T_0/20$, where $T_0$ is the period corresponding to the frequency of 2 MHz for the PZT-5H. The Transducer model was computed in the same time interval. This is a necessary condition to make compatible both classic FEM and DG models.

*Meshes*

All the meshes were "free triangular" type, except for the PMLs that were rectangular. For the Transducer Model, the mesh size was in the order of $\lambda/20$ specifically for the piezoelectric and fluid layer. For the backing layer, the size was $\lambda/40$, where it was considered as only one simple geometry domain with only one material (although inhomogeneous). The value of $\lambda$ was obtained from the lowest velocity of 1500 m/s.

In the case of Propagation Model, the same "free triangular" mesh was used with size $\lambda/1.5$. Compared to the previous case this is a great difference with a consequence of fewer degrees of freedom necessary if we compare with the Transducer Model.

*Modelling Inhomogeneous material*

The backing material model as expressed above, could be implemented in the Comsol 5.5 platform as a new two-phase composite material, given by tungsten particles randomly distributed in an epoxy resin matrix. Hence, in a 2D problem, the density and the US velocity of the composite backing are a discrete function $g(x,y)$ of the *spatial variables* $x$ and $y$, that can be expressed by the following conditional function:

$$g(x,y) = \begin{cases} \text{tungsten powder property if } f(x,y) > Thold \text{ or } f(x,y) < -Thold \\ \text{epoxy resin property if } -Thold < f(x,y) < Thold \end{cases} \quad (6)$$

Tungsten powder or epoxy-resin properties are density, Young's modulus or Poisson's ratio of each material respectively and *Thold* means threshold. This parameter defines the concentration of the tungsten power in the epoxy matrix (that means, the VFTE volume fractions). In our case with an interval value (-2,+2) as expressed later on**.** In (6), $f(x,y)$ represents a random function given by:

$$f(x,y) = \sum_{k=-N}^{N}\sum_{l=-N}^{N} a(x,y)\cos[(kx+ly)+\phi(x,y)] \quad (7)$$

Equation (7) represents a cosine spatial transform. The wave number $k$ is related to spatial frequency $f$ by $k = 2\pi f$ in the $x$ direction. There is a similar expression for the wave number $l$ and its spatial frequency in $y$ direction**.** The amplitude $a(x,y)$ is a random Gaussian distribution and the phase $\phi$ is a uniform random distribution. B. Sjodin **[19]** expressed this idea, although in our case, we do not use the filter that attenuates the high frequency components as in [19].

When using (6) in a Comsol 5.5 model, the inhomogeneous material has properties that are randomly distributed in space determined by the *Thold* level. This expression is a way to represent binary data [19]. Then the backing changes its properties in a random binary form, from tungsten to epoxy-resin according to *Thold* value.

With different *Thold* values, it is possible to change the relation VFTE between powder and epoxy-resin. Then it is possible to control the VFTE relations and as a consequence, the backing acoustic impedance model by simply changing this parameter. Hence, when the *Thold* value tends to zero, there is an increase of tungsten powder particles in the epoxy matrix. In the opposite way, when this value increases, there is more epoxy resin presented in the backing. There is a high limit value of *Thold*, where there are no more tungsten particles and only pure epoxy resin is presented in the backing. We can conclude that it is possible to control the VFTE relations and then the backing acoustic impedance model by changing the *Thold* parameter.

Fig. 3 shows the function $f(x, y)$ used in this model; it was obtained according to the expression (7) for an $N = 400$. The random function $a(x, y)$ has a Normal distribution defined for a mean = 0 and standard deviation = 1. On the other hand, the function $\phi(x, y)$ has a Uniform distribution defined for as mean = 0 and range = $+/-\pi$. The function $f(x, y)$ has limits approximately between ±2. Then, if for instance, *Thold* = 2 (or -2), the inhomogeneous material is practically pure epoxy resin (VFTE = 0). As told before, when *Thold* → 0 , there is an increment of tungsten particles and then and increasing of the VFTE relation.

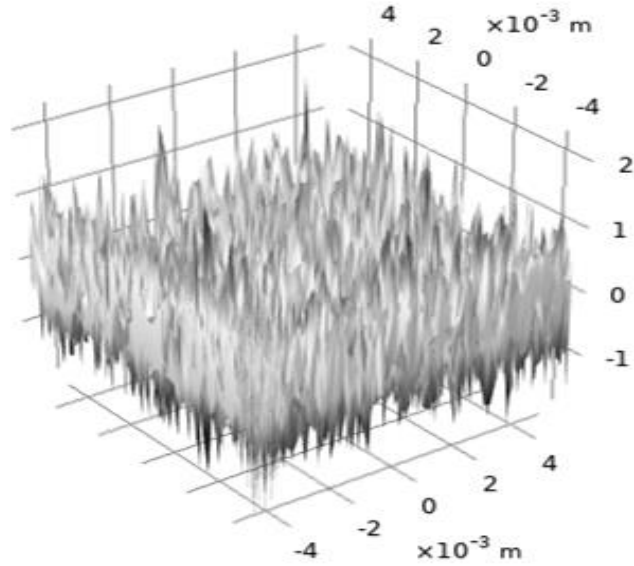

Fig. 3. Random function $f(x, y)$ used for backing material simulation, according to equation (7).

It is worth noting that the material definition expressed as (6) (and (7)) is independent of the domain meshing condition. This discrete function with its "own mesh" must be generated previously and loaded as an interpolation function (named as *inter_function*) in the Comsol Global Definitions.

As we know, the method of Galerkin uses a weak form of the physics equations. Then, using a specific base (i.e. Lagrange or Serendipity), it converts the weak form, to an algebraic equation system according to the mesh nodes [20]. Comsol has four coordinate system (Spatial, Material, Geometry and Mesh). Then in this conversion process from the weak form to the algebraic equations form, the $x, y$ coordinate values at each mesh node, take the material parameters *(density or elastic modulus)* using the *inter_function* as expressed before. On this imported function, we use the *nearest neighbor interpolation* in setting conditions of Comsol.

*Material Properties*

As active element, the PZT-5H was included with its properties from the material library in Comsol 5.5.  The mechanical damping values were chosen according to a Rayleigh model [21] with values according to Table 1.

Table 1. Rayleigh damping values for PZT-5H [17,20]

| Mass Damping parameter (alfa) [$s^{-1}$] | Stiffness damping parameter (beta) [s] |
|---|---|
| 3.0875e5 | 2.1944e-8 |

Table 2 shows a summary of the properties employed in the backing materials. In the case of Tungsten, the Comsol library was used. For epoxy resin, a new Comsol material was defined using the properties according to [7]. The Rayleigh damping values for the composite backing shown in Table 3 were taken from Ref. [10].

Table 2. Values for Tungsten and epoxy resin

| Material | Density [kg/$m^3$] | Young's modulus [Pa] | Poisson's ratio |
|---|---|---|---|
| Tungsten | 17800 | 3.6e11 | 0.28 |
| Epoxy resin | 1140 | 3.82e9 | 0.38 |

Table 3. Rayleigh damping values for backing.

| Mass Damping parameter (alfa) [$s^{-1}$] | Stiffness damping parameter (beta) [s] |
|---|---|
| 8e6 | 1.5e-9 |

It should be noted that when we include damping in piezoelectric and backing *models*, there is no more a pure elastic model. The Rayleigh damping models are good enough for pulse or transient modeling.

Finally, the fluid in the Propagation model is water and its properties were taken from the Comsol library.

*Simulations results*

The *Thold* level was considered as a parameter that varies from 0.1 to 2.5 with increments from 0.1 to 0.02, which depends on the necessary resolution according to the simulation results. Twenty-six of these *Thold* levels were used to perform numerical experiments. The simulations were carried out with Comsol 5.5 in a laptop with I7 processor 3$^{rd}$ generation and 16 GB RAM. The computational running times were four hours for the Transducer model and one hour for the Propagation models.

For the transducer model, Fig. 4 shows an example of the particle distribution in a zoom of the backing region at different *Thold* levels. In fact the Fig. 4, could correspond to density or Young Modulus diagram. At a high value of this parameter (i.e. 2.0), there is practically only one particle of tungsten in the backing area of 1 $mm^2$. As the *Thold* level decreases the density of the particles in the epoxy matrix increases. As we can see at a low *Thold* level, there is a connectivity of 0-3 that tends to 3-3 (with tungsten cluster formations) when *Thold* increases. At a value of 0.1, it is practically the opposite: a tungsten matrix with epoxy particles.

It should be also noted from Fig. 4 that for high *Thold* values, for instance 2.0 and 1.5, the tungsten diameter is less than 0.1 mm.  Using Table 2, the US velocity for the epoxy resin is approximately 2505 m/s. Then for 2-MHz frequency, the wavelength is around 1.3 mm. Hence, we can conclude that in this *Thold* range, there is a typical scattering phenomenon because the wavelength is greater than the diameter size of tungsten powder. Of course, this situation changes according to the cluster formation with the decrease of *Thold* values. The size of a tungsten particle depends on the N value in equation (7) and the mesh size of the backing ($\lambda/40$). For the model developed, as expressed above, the tungsten particles were not meshed. This is an advantage compared with previous FEM models.

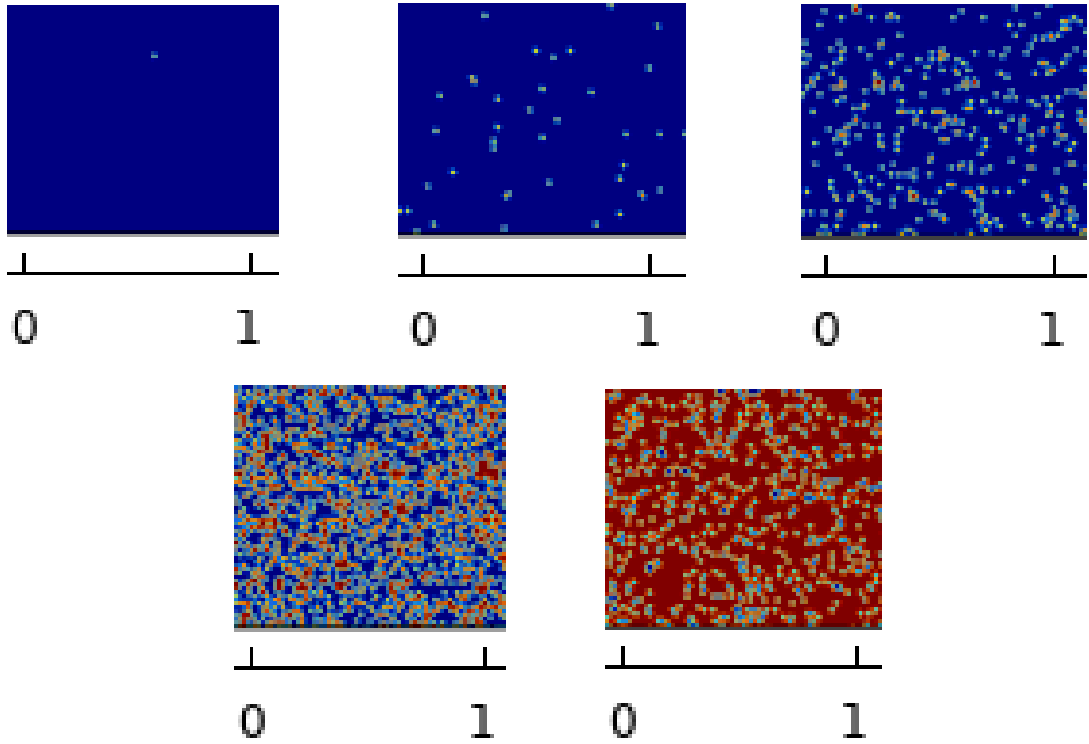

Fig. 4. Particle distribution in a zoom backing area of 1 mm$^2$ for *Thold* values of UP= 2.0, 1.5, 1.0. DOWN: 0.5, 0.1 (from left to right). The color scale covers from the maximum value (blue color) to the minimum value (red color) with a transition value at white color. The blue color corresponds to the epoxy resin while the red color corresponds to the tungsten (at the limit). The mechanical properties like density, or Young Modulus is associated at each particle color.

From the propagation model STUDY and using the same *Thold* values, Fig. 5 shows the RF signals obtained at 30 mm of the central axis. Although this point is in the near field zone, it offers the possibility of evaluating the change of the RF signal according to the *Thold* values, in a range used in medical ultrasound imaging.

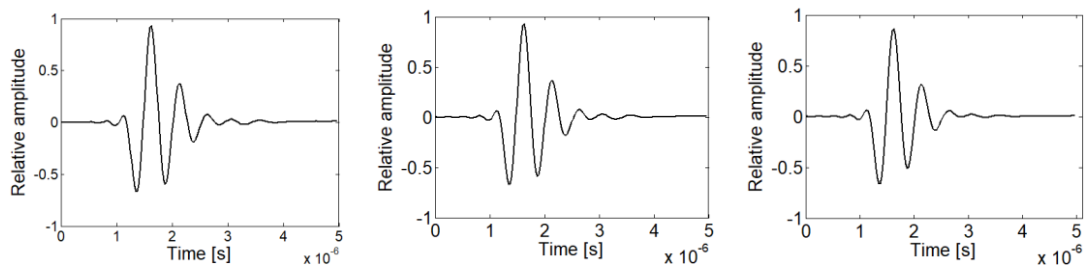

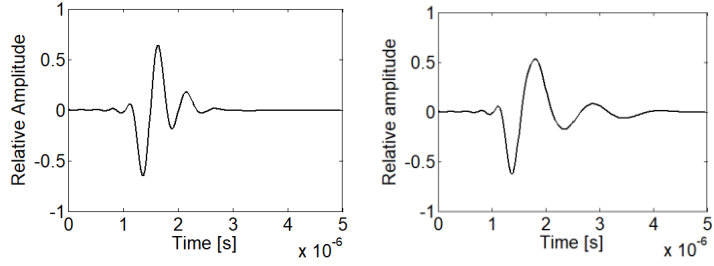

Fig. 5. RF signal obtained at 30 mm in the Propagation model domain, from *Thold* values of UP: 2.0, 1.5, 1.0; DOWN: 0.5, 0.1 (from left to right).

It is shown, in Fig. 5, only some examples of the influence in the RF signal of the *Thold* level, which determines the VFTE of the backing. Later on, we will show all the results of the central frequency and bandwidth (BW) with the 25 *Thold* levels that were performed.

In Fig. 6A, the average value of the backing density obtained using a Comsol built-in average operator is displayed as a function of *Thold*. As expected, the density decreases when increases the *Thold* value. On the other hand, Fig. 6B display the VFTE obtained from (5) assuming that $\rho_B$ is represented by the average density, obtained from the Comsol operator. These figures show the new idea where several backing acoustic impedances can be simulated through the *Thold* value, which determines the VFTE. From this, the transducer performance can be evaluated on the final RF signal emitted in the medium. Hence, it is not necessary to obtain the real acoustic impedance of the backing to describe the transducer radiation performance. Only the VFTE through the *Thold* value is necessary to obtain the final RF transducer response.

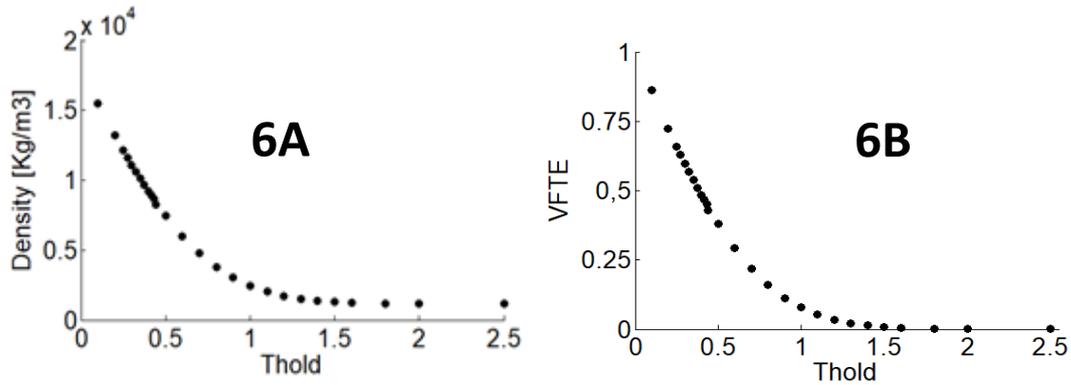

Fig. 6. (6A) Average density vs *Thold* values. (6B) VFTE vs *Thold* values.

From the RF obtained for all *Thold* values considered in the simulation experiments, it was possible to calculate the bandwidth (BW) and the central frequency $f_0$ using FFT on Matlab Platform. These parameters are given by the following expression [22]:

$$BW = (f_2 - f_1) \qquad (7)$$

Where $f_1$ and $f_2$ are the frequencies at 3 dB below the peak frequency value. The central frequency $f_0$ is defined as [22]:

$$f_0 = \sqrt{f_1 * f_2} \qquad (8)$$

Fig 7 shows the behavior of the central frequency obtained from (8) with several VFTE values. As expected at low values, the emitted pulse has a central frequency close to 1.8 MHz, which corresponds to the original piezoelectric frequency of 2 MHz but affected by this "light backing" region. When the VFTE value is close to 0.5, there is an inflexion point in the curve. A region of "heavy backing" appears after this point and the frequency of the pulse is tending to the half of the piezoelectric value. In this situation, the nodal plane of the piezoelectric element is no longer situated in the central piezoelectric plane, as it should be in a free piezo-element. For this case, this plane is displaced close to the boundary between the piezoelectric elements and the backing. This is like a piezoelectric image inside the backing.

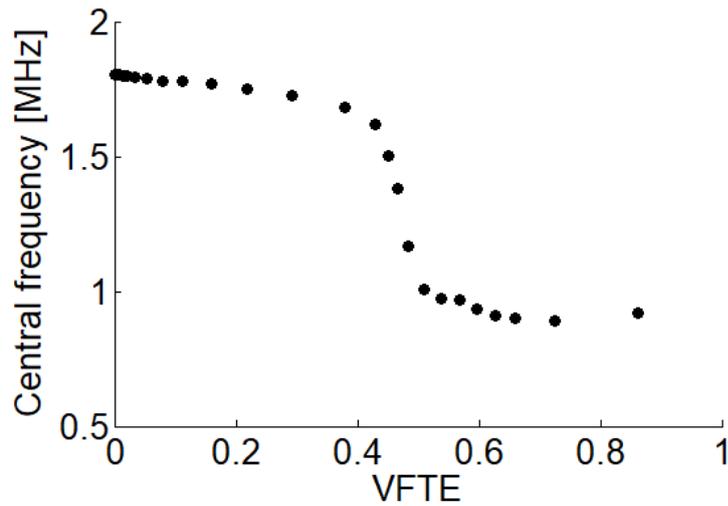

Fig. 7. Pulse central frequency relation with the VFTE.

Fig. 8 shows the bandwidth as a function of the VFTE. In this example, the maximum occurs when the VTFE value is approximately 0.538. This means that the volume of tungsten necessary to obtain the optimal axial resolution is 53.8% of the epoxy resin volume. Condition (3) predicts an acoustic impedance around 20 MRayls assuming a value of 30 MRayls for PZT-5H. Although our model does not need the acoustic impedance of the backing, we can evaluate approximately this parameter from the graphics of the impedance of metal-loaded backing in Ref [2]. In this reference, the acoustic impedance value of 20 MRayls is obtained around 55% of the tungsten volume in epoxy resin. This is close to our prediction for the maximum bandwidth and, of course, close to the Desilets condition.

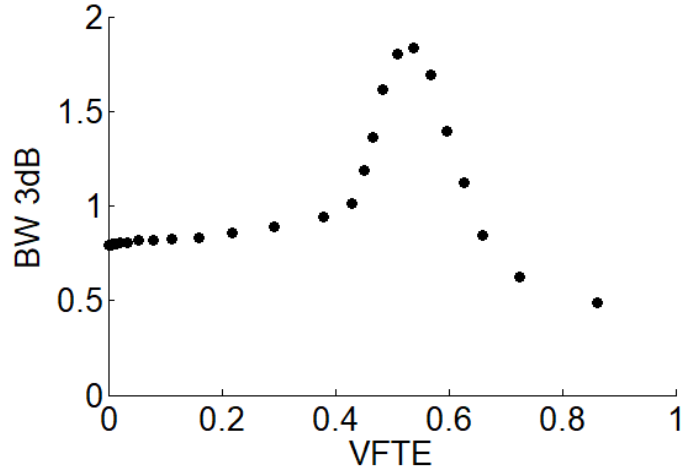

Fig. 8. Bandwidth at 3dB relation with the VFTE.

Our model was calculated in 2D to avoid memory issues associated with 3D models. It is important to note that Desilets obtained his results using a KLM circuit. The KLM circuit simulation assumes a "piston like" vibration of the transducer face. However, FEM model is closer to a real situation where the face vibration of the transducer is no more a piston-like as expressed by Guo and Cawley [23]. In any case, both results KLM and FEM are close to each other. We can conclude that the Desilets expression (3) is also a good condition of maximum BW in "non- piston like" models.

In our model, we avoid the necessity of using the acoustic impedance value in the backing section modeling. As expressed before, the fundamental idea in our work is to predict the RF signal finally emitted in function of the volume fraction of Tungsten in epoxy (or VFTE). Hence, the FEM model proposed here could represent a good tool for predicting the real signal emitted in backed transducers according to many composite arrangements. The model developed, using inhomogeneous material with the combination of CG and DG, shows that it is possible to simulate transducers with dimensions close to those encountered in the commercial transducers.

Finally, in Fig 9, an example of two signals obtained from transducers with backings of tungsten and alumina respectively at the same volume percentage is shown. One can observe the difference in the pulse length that is longer for the alumina composite than for the tungsten composite (both in an epoxy resin matrix), as it corresponds to the low acoustic impedance of the alumina material.

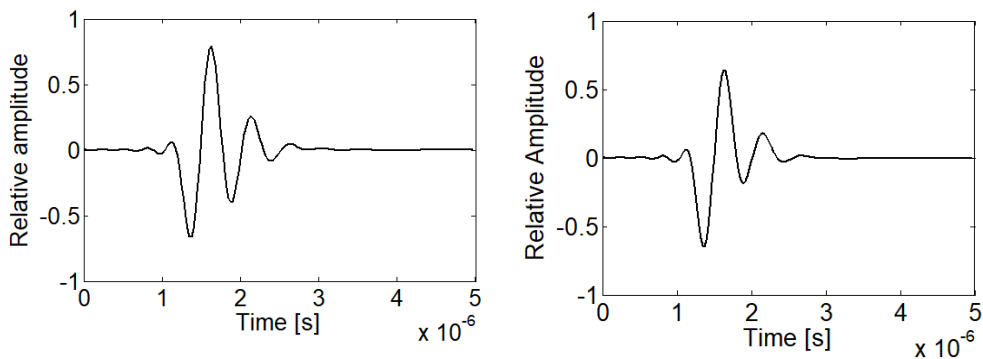

Fig. 9. RF signals obtained from a transducer with a backing of alumina powder (left) and tungsten (right) powder in epoxy matrix with *Thold* = 0.5. As expected, the alumina-backing signal has more cycles than the similar one with tungsten.

*Conclusions*

A 2D FEM model for simulating, in transient regime, the ultrasonic transducer performance with composite backing is presented. This is proposed assuming a classic circular piezoelectric transducer or the rectangular array element. The backing was modeled using one single geometric domain with only one inhomogeneous material. This overcomes the complex situation of two-phased FEM models, where the classical complex geometry defined by tungsten particles plus epoxy resin matrix should be used. The transducer performance was evaluated from the RF simulation of the pulsed propagation in a fluid as a function of the volume relation of tungsten inside an epoxy resin matrix.

The model developed could serve as a good tool in transducer engineering, where it is necessary to design ultrasonic transducers according to the broadband RF signal finally emitted. Although the model is formed with a piezoelectric vibrator and only one composite backing; it is possible to include in the future several coupling layers with the propagation medium.

In this work, we expose a composite backing with only two components. However, it could be possible to improve the model with three components, for instance an epoxy resin matrix mixed with two different powders. The used idea here of inhomogeneous material is good enough for all these purposes, because the composite backing FEM model was considered as only one material. Studies with these improvements, including a 3D model could be done in future works.

*Acknowledge*

We would like to thank to the Ditecrod Network of the CYTED Program (www.cyted.org) for the helping in the collaboration among the groups that participated in this work. We also thank to CSIC (Spain) and the Brazilian agencies CAPES, CNPq and FAPERJ for the financial support.